\documentclass[12pt,preprint]{aastex}

\slugcomment{Submitted to AJ, June 2003}

\shorttitle{The morphological Luminosity Functions of A868}
\shortauthors{Driver et al.}

\begin{document}


\title{The morphological decomposition of Abell 868\footnote{Based on
observations made with the NASA/ESA Hubble Space Telescope, obtained
at the Space Telescope Science Institute, which is operated by the
Association of Universities for Research in Astronomy, Inc., under
NASA contract NAS 5-26555. These observations are associated with
program \#8203.}}


\author{S.P.Driver}
\affil{The Research School of Astronomy and Astrophysics, The Australian 
National University, Weston Creek, ACT 2611, AUSTRALIA}
\email{spd@mso.anu.edu.au}

\author{S.C.Odewahn, L.Echevarria, S.H.Cohen, R.A.Windhorst}
\affil{School of Physics and Astronomy, Arizona State University, Tempe, 
AZ 85287-1504, USA}

\author{S.Phillipps}
\affil{Department of Physics, University of Bristol, Bristol, BS8 1TL, UK}

\and

\author{W.J.Couch}
\affil{School of Physics and Astronomy, University of New South Wales,
Sydney, NSW 2052, AUSTRALIA}



\begin{abstract}
We report on the morphological luminosity functions (LFs) and radial
profiles derived for the galaxy population within the rich cluster
Abell 868 ($z=0.153$) based purely on {\it Hubble Space Telescope}
Imaging in $F606W$.  We recover Schechter functions ($-24.0 <
M_{F606W}-5\mbox{log}h_{0.65} < -16.0$) within a $0.65 h_{0.65}$ Mpc
radius for early(E/S0)-, mid(Sabc)- and late(Sd/Irr)- type galaxies
of:

~

$M^*_{All} - 5 \log_{10} h_{0.65} = -22.4^{+0.6}_{-0.6}, 
~\alpha_{All} = -1.27^{+0.2}_{-0.2}$

$M^*_{E/S0} - 5 \log_{10} h_{0.65} = -21.6^{+0.6}_{-0.6}, 
~\alpha_{E/S0} = -0.5^{+0.2}_{-0.3}$

$M^*_{Sabc} - 5 \log_{10} h_{0.65} = -21.3^{+1.0}_{-0.9}, 
~\alpha_{Sabc} = -1.2^{+0.2}_{-0.2}$

$M^*_{Sd/Irr} - 5 \log_{10} h_{0.65} = -17.4^{+0.7}_{-0.7}, 
~\alpha_{Sd/Irr} = -1.4^{+0.6}_{-0.5}$.

~

\noindent
The early-, mid- and late- types are all consistent with the recent
field morphological LFs based on recent analysis of the Sloan Digital
Sky Survey --- Early Data Release (SDSS-EDR; \citealp{nakamura03}).
From a detailed error analysis, including clustering of the background
population, we note that improved statistics can only come from
combining data from many clusters.

~

\noindent
We also examine the luminosity-density and number-density profiles as
a function of morphology and draw the following conclusions: (1) The
galaxies responsible for the steep faint-end slope are predominantly
of late-type morphology, (2) The cluster core is dominated by
elliptical galaxies, (3) The core is devoid of late-types systems, (4)
The luminosity-density as a function of morphological type is skewed
towards early-types when compared to the field, (5) Up to half of the
elliptical galaxies may have formed from the spiral population through
core disk-destruction process(es).

~

\noindent
We believe the most plausible explanation is the conventional one that
late-types are destroyed during transit through the cluster core and
that mid-types are converted into early-types through a similar
process, which destroys the outer disk and results in a more tightly
bound population of core ellipticals.
\end{abstract}




\section{Introduction}
The overall luminosity distribution of galaxies in any environment is
the traditional tool for describing the galaxy population (see
\citealp{bst88}, BST). However, while it categorises the
number-density as a function of absolute magnitude it provides no
information on the morphology, structure, spectra or star-formation
rates of the contributing galaxies. While studies may show that the
luminosity function (LF) of the field, groups and rich clusters are
comparable at bright magnitudes (see for example \citealp{dp03} and
\citealp{christlein03}), this is by no means conclusive proof that the
entire galaxy population and characteristics are identical. Indeed the
morphology-density (\citealp{dressler80,dressler97}) and the dwarf
population-density \citep{phillipps98} relations clearly tell us that
local galaxy density is important and that luminous elliptical
galaxies prefer clustered environments and low luminosity irregular
galaxies field environments. In short, a single luminosity
distribution may bypass exactly the information that is required to
decipher the subtleties of the environmental dependency of galaxy
evolution.

In addition, recent measurements of the LFs in rich clusters have led
to inconsistent conclusions as to whether there is a universal LF (see
for example \citealp{trent98}) or a dwarf population-density relation
\citep{phillipps98}. In a study of 7 Abell clusters \citet{driver98a},
using a statistical background subtraction method, found significant
variation in the faint-end slopes whereby low density clusters exhibit
steeper slopes (or higher dwarf-to-giant ratios). The same result was
independently found for a separate sample of 35 clusters by
\citet{lopez97}\footnote{It is worth noting that both studies used a
fixed field-of-view size, limited by the respective detectors, and
hence representing progressively larger physical extents for higher
redshift clusters.}.  However, both methods rely on a statistical
subtraction of the background population which, although rigorously
tested in \citet{driver98b}, has been criticized by \citet{valotto01}
as being susceptible to cosmic variance along the line of sight ---
although, it is difficult to understand how cosmic variance can lead
to the relatively clean relation between luminosity and local density
seen by \citet{phillipps98} and the smooth radial increase in
dwarf-to-giant ratios seen in A2554 \citep{smith00} and A2218
\citep{pracy03}. More recently \citet{bark03} report a general trend
of an increase in faint-end slope with cluster radius from $\alpha =
-1.81$ to $\alpha=-2.07$ for a sample of 17 nearby clusters.  For very
local clusters where cluster membership can be ascertained more
easily, such as Virgo and Coma, \cite{trent02} summarise the state-of
play and argue for a universal LF (see also review by
\citealp{driver03} and references therein). \citet{trent01} however
argue the opposite noting the significant difference in dwarf-to-giant
ratio between Virgo and Ursa Major. Some part of this confusion most
likely comes about from the apparent different clustering of the two
dwarf populations. For example \citet{sand85} found a generally
centrally concentrated distribution of dwarf ellipticals in Virgo,
whereas \citet{sab03} report to the contrary a significant steepening
in the luminosity function faint-end slope with cluster-centric
radius, also in Virgo, due to low surface brightness dwarf irregulars.
In the Coma cluster \citet{thompson93} identify three dwarf
populations (dIs, dEs and dSphs) each with distinct clustering
signatures.

Taken together, the sparse information contained within a single LF
and the contradictions in the literature, it seems necessary to
deconstruct the LF further, incorporating morphological/structural
and/or color information in the analysis. It is also worth noting
that some component of the confusion may arise from radial
dependencies and the specific areal extent over which the cluster has
been surveyed --- particularly if the above radial trends seen in
Virgo are confirmed as universal. To this end we have embarked upon a
detailed observational program, including space-based optical and
X-ray observations, and ground-based narrow-band imaging, of the rich
cluster A868. In this paper we focus purely on the morphological
aspects based upon a 12 orbit {\it Wide Field Planetary Camera 2}
mosaic of the cluster A868. In particular we are interested in the
suggestion that there may exist a universal LF for each morphological
type (BST) and that only the relative normalisation changes with
environment. Analysis of the two-degree field galaxy redshift survey
by \citet{dp03} find that although the overall luminosity distribution
is invariant between the field and cluster composite, differences do
arise when subdivided according to {\it spectral}
type. \citet{christlein03} confirm this result based on their
independent spectral study of the population in and around 6 low
redshift clusters.  These latter results, based on spectral
classifications, generally supports the developing notion that
star-formation is quenched in the infalling galaxy population
(\citealp{lewis02}, \citealp{gomez03}, see also review by
\citealp{bower03}), unfortunately spectral classifications cannot
address whether the population has {\it physically} changed as well.

The cluster A868 itself, is unremarkable, except that it formed part
of a cluster population study by \citet{driver98a}, in which a high
dwarf-to-giant ratio was found. The primary purpose of these HST data
were to study the morphologies and structural properties of the giants
and dwarfs, and in particular to identify the nature of the population
responsible for the apparently steep LF upturn at the faint-end. An
initial attempt in this regard, using ground-based data, was made by
\citet{boyce01}. They concluded that the population responsible for
the faint upturn could be subdivided into three categories: a
contaminating population of background high-redshift ellipticals, an
overdensity (relative to the giants) of dwarf ellipticals, and an
overdensity of dwarf irregulars. The type classification was made on
the basis of color.  \citet{boyce01} noted that when the
population of contaminating background galaxies was removed, the
overall LF still showed a distinct upturn ($\alpha = -1.22$) and a
generally high overdensity of dwarf galaxies. From the colors it was
concluded that the main component of this population was blue and
therefore presumed to consist of dwarf Irregular galaxies.
Furthermore \citet{boyce01} argued that the core was devoid of dIrrs
which were mostly destroyed via processes such as galaxy harassment
\citep{moore98}, thus accounting for the increase in the luminosity
function faint-end slope from the centre outwards \citep{driver98a}.

The plan of this paper is as follows: In section 2 we summarise the
observations, reduction and analysis of the {\it Hubble Space
Telescope} images. In section 3 we describe and validate the
morphological classification process, and in section 4 we describe the
appropriate error analysis incorporating the clustering signature of
the background population.  In section 5 we show the overall and
morphological luminosity distributions, determined via statistical
background subtraction, and compare them to recent field estimates to
test BSTs hypothesis. In section 6 we investigate the radial
distribution in terms of the luminosity- and number- density profile
of each morphological type and conclude in section 7. We adopt
$H_{o}=65$ km/s/Mpc, $\Omega_{M}=0.3$ and $\Omega_{\Lambda}=0.7$
throughout, this results in a distance modulus to A868 of $39.47$ mags
(excluding K-correction).

\section{Data Acquisition, Reduction and Analysis}
A868 formed part of a cluster population study by \citet{driver98a},
in which a high dwarf-to-giant ratio was found (see also
\citealp{boyce01}). To pursue this further 12 orbits were allocated in
Cycle 8 with the {\it Wide Field Planetary Camera 2} onboard the {\it
Hubble Space Telescope} (HST\footnote{Based on observations made with
the NASA/ESA Hubble Space Telescope. STScI is operated by the
Association of Universities for Research in Astronomy, Inc. under NASA
contact NAS 5-26555.}) to obtain a six-pointing F606W mosaic of the
cluster. A868 lies at coordinates $\alpha_{J2000.0} = 09h45m26.43s,
\delta_{J2000.0} = -08^o39^{'}06.7^{''}, z=0.153$
\citep{strubble99}. The cluster has an Abell richness class 3, and is
of Bautz-Morgan type II-III (see \citealp{driver98a} and
\citealp{boyce01} for the earlier work on A868).

The data comprise 24 individual exposures of 1100s, each targeted at
six individual and marginally overlapping pointings (see Fig.~1).  The
data were combined using a pixel clipping algorithm based on local sky
statistics developed for use with {\sc WFPC2} images in the {\sc
lmorpho} package \citep{odewahn02}. Extensive tests were made
comparing the photometry derived from such stacks, to those derived
from the {\sc drizzle} algorithm \citep{fruchter02} with no
appreciable systematic difference found. The {\sc lmorpho} stacks,
produced in a more straight-forward fashion, and free of problems
associated with correlated pixel noise, were adopted for further
use. The final pixel scale is 0.0996 arcsec/pixel and the full mosaic
field covers an area of 0.007545 sq. degrees.  Fig.~2 shows the WFPC2
chip containing the cluster core, showing the dominant cD and D
galaxies, and evidence for strong gravitational lensing.  The
photometric zeropoint for each mosaic was 30.443, as taken from
\citet{holtzmann95}, placing the photometry onto the Vega system.
Initial object source catalogs were derived with {\sc se}xtractor
\citep{bertin96} using a 2$\sigma$ sky level threshold (per pixel) and
a minimum isophotal area of 5 pixels. A {\sc gui}-based image editor
in the {\sc lmorpho} package was used to visually inspect image
segmentation over the field and edit obvious problems. Image postage
stamps were prepared for each detected source and the {\sc galphot}
package in {\sc lmorpho} was used to perform automated galaxy surface
photometry. This package incorporates information about nearby
cataloged sources and performs modest corrections designed to decrease
photometric degradation from field crowding. The {\sc lmorpho} catalog
for 1616 valid objects in A868 contained a variety of image structural
parameters as well as total magnitudes and quartile radii (including
the effective radius) --- for full details of the inner workings of
this software package see \citet{odewahn02}. Note that final
magnitudes are extinction corrected using \citet{schlegel} dust maps.
Briefly an initial isophotal magnitude within an elliptical aperture
is measured and the data is corrected to total based upon the
extrapolated profile fit. In most cases this provides an excellent
approximation to the total magnitude and is ideal for crowded
sight-lines such as A868. However its well known that for anomalous
and/or flat profile objects the isophotal correction can become
unrealistically large. As a check of the isophotal corrections we show
the isophotal versus total magnitudes for the full A868 galaxy
population (see Fig.~3). Clearly a small fraction of objects do indeed
have unrealistic isophotal corrections. We hence adopt a cap to the
isophotal correction shown as the dotted line. This is a simple
power-law fitted to the lower bound of the brighter data (note that
not surprisingly the cap is only required for the late-types,
triangles on Fig. 4, which exhibit non-standard profile shapes). The
expression for the isophotal cap is given by: $m_{Total} \leq m_{Iso}
+ 0.25 - 0.0055 (m_{Iso}-16)^{2.5}$

Finally Fig.~4 shows the apparent bivariate brightness distribution,
this highlights that stars and galaxies are well separated to
$m_{F606W} \leq 24.0$ mag, and that the bulk of the galaxy population
lies above the surface brightness detection isophote to the same
limit.  From Fig. 4 it is also apparent that earlier-types are of
higher effective surface brightness in line with conventional wisdom.


\subsection{Reference field counts}
In order to determine the contribution to the A868 galaxy counts from
the field, we performed an identical reduction and analysis on the
Hubble Deep Field North (HDFN), Hubble Deep Field South (HDFS) and the
deep field {\sc 53W002} \citep{driver95, wind98} --- all observed in
F606W, covering $\sim 0.0011$ sq deg, and calibrated onto the same
photometric system as A868 (see \citealp{cohen03} for further details
of these specific fields). However these three deep fields only
provide reference counts at faint magnitudes ($m_{F606W} > 21$
mag). To provide reference counts at brighter magnitudes we adopt the
Millennium Galaxy Catalogue (MGC; \citealp{liske03, cross03}) and
convert the MGC photometry from $B_{MGC}$ to F606W.  This is achieved
by convolving the {\it Isaac Newton Telescope's} KPNO $B$ and the {\it
Hubble Space Telescope's} F606W filter+instrument transmission
functions with the mean zero redshift cosmic spectrum from the 2dF
galaxy redshift survey \citep{baldry02}, after dividing out the
equivalent flux calibrated spectrum for Vega (see for example
\citealp{sung00}). This resulted in a transformation of:
$(B_{MGC}-F606W)_{Vega}=+1.06$. Although the MGC counts extend to
$B=24$ mag, the color transformation above will only be appropriate
for non-cosmological distances, i.e., $B \leq 18.25$ mag.

\section{Galaxy Classification}
Object classification for the A868, HDFN, HDFS, 53W002 fields were
performed using an Artificial Neural Network (ANN), as described in
\citet{odewahn96}. Briefly, the ANNs were initially trained on a
sample classified by eye, drawn from a variety of datasets including
the HST BBpar \citep{cohen03} and RC3 catalogues \citep{devauc95}.
The ANNs take as input parameters a set of structural measurements for
each image (seven isophotal areas and a seeing/PSF measurement) and
output a classification onto the 16 step de Vaucouleurs' t-type system
(see \citealp{devauc95}) with an additional step added for
stars. Stars are defined as t-type=12, early-types (E/S0) as $-6.0
\le$ t-type $\le 0.0$, mid-types (Sabc) as $0.0 <$ t-type $\le 6.0$,
and late-types (Sd/Irr) as $6.0 <$ t-type $\le 10.0$. An error is
allocated to each classification based upon the dispersion amongst
five independently trained ANNs. As a check of the classification
accuracy we visually inspected all objects brighter than $m_{F606W} <
24.0$ mag. In 80 out of the 663 cases a visual override was
necessary. The majority of these were due to entangled isophotes
(i.e., crowding) which is known to cause some problems with ANN
classifications. Table 1 sumarises the overrides and no obvious
classification bias is apparent. We also note that three of these
errors were the A868 central cD and two D galaxies which were all
erroneously classified as Sabcs. As no cD or D galaxies were included
in the ANN training sets, it is understandable that the giant bulge
surrounded by a low surface brightness halo could readily be confused
with a mid-type spiral. Excluding these three specific objects,
thereby gives an unchecked ANN classification accuracy of $\sim 90$
per cent.  Postage stamp images for randomly selected galaxies are
shown in Fig. 5, ordered by type and apparent magnitude.

For the ground-based MGC data all galaxies brighter than $B \leq
18.25$ were classified by eye (SCO) to provide fully
consistent\footnote{This process produces fully consistent counts as
the ANNs were trained on data classified by SCO.}  bright magnitude
reference counts.

\section{Error Analysis}
Prior to field subtraction it is first worth making careful
consideration of the error budget, particularly in light of concerns
raised by \citet{valotto01} that many of the steep faint-ends observed
in clusters, are due to the clustering signature of the background
field population. This has some justification as the error analysis
involved when subtracting reference counts from cluster counts has
often been overlooked (for example in \citealp{driver94}). Here we
intend to extend the normal analysis to now incorporate this
additional error component.

~

\noindent
In this particular analysis there are five components to the error
budget: Counting errors in; the reference counts ($\sigma_{R}$), the
field counts in the cluster sight-line ($\sigma_{F}$), and the cluster
population itself ($\sigma_{C}$); along with the clustering error in
the two sets of field counts ($\psi_{R}$ and $\psi_{F}$). Note that we
separate out the two counting errors in the cluster sight-line as in
reality there are two distinct superimposed populations (field, F and
cluster, C). For all three counting errors we adopt the usual
assumption of $\sqrt{n}$ statistics for the associated error ({\it
i.e.,} Poisson statistics).  For the clustering error we start from
the prescription given in \citet{peebles80} [Eqn. 45.6] which provides
an expression for the total variance in cell-to-cell counts for a
randomly placed cell as:
\begin{equation}
<(N-n\Omega)^{2}> = n\Omega + n^{2} 
\int d\Omega_{1}d\Omega_{2}\omega(\theta_{12})
\end{equation}
here N is defined as the counts in a given cell (i.e., per
field-of-view, $\Omega$), n is the global mean count per sq degree and
$\theta_{12}$ is the separation between the solid angle elements
$d\Omega_{1}$ and $d\Omega_{2}$.  In this expression the first term
represents the Poisson error ($\sigma$) and the second the clustering
error ($\psi$), i.e.,
\begin{eqnarray}
\sigma^{2} = & n\Omega \\
\psi^{2} = & n^{2} \int d\Omega_{1}d\Omega_{2}\omega(\theta_{12}) \\
\approx & n^{2} \theta^{4} \omega(\frac{\sqrt{2}\theta}{3}) \\
= & n^{2} A_{w} \theta^{3.2}(\frac{\sqrt{2}}{3})^{-0.8}
\end{eqnarray}
The above simple approximation for $\psi$ uses the mean
separation between points in a square of side $\theta$
\citep{phillipps85} and the standard expression for the angular
correlation function of $\omega(\theta) = A_{w} \theta^{-0.8}$. Replacing
$n\Omega$ with $N(m)$ (the number-counts for the specified
field-of-view) and $A_{w}$ with $A_{w}(m)$ yields the variances from
the clustering error for any field size ($\Omega$ or $\theta^{2}$) and
magnitude interval ($m$). Observationally we find \citep{roche98}
that:
\begin{eqnarray}
A_{w}(m) = & 10^{-0.235m_{r} + 2.6}
\end{eqnarray}
Hence by combining Eqns~5 \& 6 and adopting $(F606W-R)=0.2$---$0.6$ we get a
final approximation for $\psi$ of:
\begin{equation}
\psi^{2} \approx 1.83 N(m_{F606W})^2 10^{(-0.235 m_{F606W} + 2.7)} \Omega^{-0.4}
\end{equation}
Assuming $\omega$ remains a power law out to the size of the field.
Here $N(m_{F606W})$ are the galaxy counts per 0.5 mag for the
specified field of view, $\Omega$, which is given in sq degrees.

The five errors identified above can now be written down
explicitly as follows:
\begin{eqnarray}
\sigma_R = & \frac{\sqrt{N_R}{\Omega_C}}{3 \Omega_R} \\
\psi_R = & \sqrt{3 (1.83 (N_R/3)^{2} 10^{(-0.235 m+2.7)} \Omega_{R}^{-0.4})} \frac{\Omega_C}{\Omega_R} \\
\sigma_C = & \sqrt{N_{C}} \\
\sigma_F = & \sqrt{N_F} \\
\psi_F = & \sqrt{(1.83 (N_F)^2 10^{(-0.235 m+2.7)}\Omega_{C}^{-0.4})}
\end{eqnarray}
Where $N_R, N_F$ and $N_C$ are the number-counts for the combined
reference fields, the field population in the A868 sight-line and the
number-counts of the cluster population respectively, and $\Omega_R$
and $\Omega_C$ are the field-of-views of the 3 individual reference
fields (0.0011 sq deg) and the cluster field-of-view (0.007545 sq deg)
respectively.  Where appropriate these errors, or their adaptations,
are combined in quadrature and used throughout all further analysis
steps.

\section{The Morphological Luminosity Distributions of A868}
The overall and morphological galaxy number-counts for the full A868
mosaic and the combined reference fields scaled to the same area are
shown on Fig.~6.  Note that the A868 total counts lie above the
reference field counts until $m_{F606W} \approx 24.25$ (and for each
class until $m_{F606W}^{E/S0} \approx 25.25$, $m_{F606W}^{Sabc}
\approx 24.75$, $m_{F606W}^{Sd/Irr} \approx 24.0$), at which point the
A868 counts drop sharply indicating the approximate completeness
limit(s) of the A868 data (see also Fig. 4). We hereby adopt
$m_{F606W} \approx 24$ mag as the completeness limit (equivalent to
$M_{F606W} = -16$ mag) and 0.75 mag brighter than the apparent
completeness limit. The reference counts, obtained from the two Hubble
Deep Fields and 53W002, extend substantially deeper than the A868
counts but provide no available data at bright magnitudes ($m_{F606W}
< 21$ mag). To circumvent this we add in the MGC bright counts after
transposing from B to F606W as discussed in section 3.  To provide
continuous coverage over the full magnitude range we now elect to
represent the field counts by a second order polynomial
fit\footnote{The fit is a least squares fit to the data with the
errors given by Eqns. 8 and 9.} to the combined reference field
data. As well as providing continuous coverage this has the additional
advantage of smoothing the reference data to remove unwanted structure
from the 3 contributing fields. The field data used and the resulting
fits are shown in Tables 2 \& 3 respectively. Note that the data were
only fitted over the magnitude range $15.75 < m_{F606W} < 24.25$
although additional data are shown in Table 2 for completeness. The
smoothing of the counts does not reduce the associate errors but
redistributes it over the specified magnitude range.

Subtracting the smoothed reference field counts from the A868 counts
for each population yields a direct statistical representation of the
morphological luminosity distribution for the cluster (adopting a
universal Sab K-correction of $0.20$ mags), as shown on Fig.~7 and
tabulated as Table 4. Also shown on Fig.~7 (upper left, dotted line)
is the 2dFGRS composite cluster luminosity function (LF) as derived by
\cite{dp03}, shifted to the F606W bandpass. This gives a formally
acceptable fit to the cluster. The open squares show the previous and
deeper ground-based $R$-band data which agrees well within the
errors. Given that the background subtraction is derived from an
entirely different region of sky to the earlier work (see
\citealp{driver98a}) this provides a further indication that the steep
faint-end slope seen in A868 is a robust result. Of course one might
argue that the A868 sight-line could be contaminated by a more distant
cluster, although this would boost the faint elliptical counts/LF
which is not seen. Fig.~7 shows the LFs of ellipticals (E/S0s, upper
right), spirals (Sabcs, lower left) and irregulars (Sd/Irrs, lower
right). Morphological K-corrections of $K(E/S0)=0.25$, $K(Sabc)=0.20$,
$K(Sd/Irr)=0.11$ were calculated for the F606W filter combined with
the 15Gyr evolved E, Sa and Sc model spectra of \citet{bianca97}.  The
formal $1, 2, \&~ 3 \sigma$ error ellipses for the Schechter function
fits, based on the $\chi^{2}$-minimisation of the standard Schechter
LF, are shown as Fig.~7.  The results and formal $1\sigma$ errors are
also tabulated in Table 5.  For Fig.~7 (upper left, all types) the
solid line shows the sum of the three individually derived
morphological LFs showing interesting structure consistent with recent
reports of an upturn at fainter magnitudes (e.g.  A0963,
\citealp{driver94}) and/or a dip at intermediate mags (e.g. Coma,
\citealp{trent98}). If each morphological class has a universal LF, as
has been suggested \citep{bst88}, this dip then naturally arises as
the morphological mix changes (as required by the morphology-density
relation, \citealp{dressler97}). The errorbars shown in Fig.~7 and the
resulting error contours shown on Fig.~8 include the five error
components discussed in section 4. It is worthwhile assessing which of
these error components dominate the error budget. Fig.~9 shows the
total and individual error components involved in this analysis. From
this figure we can see that the dominant error at bright magnitudes
comes from the number of cluster members, whereas at faint magnitudes
the dominant error typical comes from the clustering of the background
population in the cluster sight-line. One interesting point to note is
that a full blown spectroscopic study would fail to reach the faint
magnitudes probed here, and of course be unable to improve the
statistics at bright magnitudes. In fact a spectroscopic study is more
likely to lead to additional uncertainty due to completeness
issues. Further improvement can only come from the combination of
extensive deep {\it imaging} data for a large sample of combined
cluster data. Nevertheless it is clear from Fig. 7 that the steep
faint-end seen in A868 is almost entirely dominated by late-types with
some contribution from mid-types in general agreement with the
findings of \citet{boyce01}.

\subsection{Comparisons with the Field}
Unfortunately while field morphological LFs exist, no comparison is
sensible unless an identical morphological classification methodology
has been applied. However as a general result, morphological field
studies typically find $\alpha > -1$ for early-types, $\alpha \approx
-1$ for mid-types and $\alpha < -1$ for late-types (see for example
SSRS2s morphological LFs, \citealp{marzke98}; 2dFGRSs spectral LFs,
\citealp{dp03}, \citealp{madgwick02}; and SDSS-EDRs morphological LFs,
\citealp{nakamura03}). We compare the recent results, from the
SDSS-EDR, who classify 1500 galaxies onto a similar but not identical
morphological system in $r'$. To adapt Nakamura et al's numbers to
provide consistent morphological LFS we added 2/3s of their Sbc-Sd
class to their S0a-Sb class and 1/3 of their Sbc-Sd class to their Im
class and rederived the Schechter function parameters. We also derive
$(F606W-r')=r'+1.35 (B-V) - 0.95$ (from \citealp{fukugita96};
\citealp{liske03}; and our estimate of $(B_{MGC}-F606W)$ from section
2.1) and adopt $(B-V)_{E/S0}=0.9$, $(B-V)_{Sabc} = 0.7$ and
$(B-V)_{Sd/Irr} = 0.5$ (see \citealp{driver94}). This gives the
following morphological field LFs:

$M^*_{E/S0} = -21.42, \alpha_{E/S0} = -0.8,$ 

$M^*_{Sabc} = -21.35, \alpha_{Sabc} = -1.1,$ 

$M^*_{Sd/Irr} = -21.65, \alpha_{Sd/Irr} = -1.9$

The location of the morphological field LFs are shown on Fig.~8 as
solid symbols with errorbars. We see that the E/S0, Sabc and Sd/Irr
field and cluster LFs are all consistent at the $1 \sigma$-level (and
in qualitative agreement the field and cluster LFs of \cite{bst88}).
Clearly though the errors dominate and many clusters must be studied
in a combined analysis before the universality of morphological
luminosity functions can be confirmed or refuted. Given the extensive
SDSS-EDR database and the incoming ACS cluster data this is likely to
be established in the near future and the current results should be
taken as indicative that the morphological LFs are not widely variant
between cluster and field environments.

\section{The Morphological Radial Distributions of A868}
We now subdivide the mosaic into five radial intervals of $0.75'$ (130
kpc) around the dominant cD, and calculate the contribution of each
morphological class to the luminosity- and number- density within the
range $15.9 < m_{F606W} < 23.9$ mag (equivalent to $-24 < M_{F606W} <
-16$ mag).  To achieve this we build a map of the mosaic to calculate
the relevant active fields-of-view, within each annulus, and use the
expressions given in Table 3 to subtract off the appropriate field
component.  Fig.~10 (upper) shows the radial dependency of the
luminosity-density, $j$, for each type in arbitrary units and Fig.~10
(lower) the number-density.  Whereas the former is skewed towards
brighter systems (which dominate the luminosity-density) the latter is
skewed towards fainter systems (at least for mid- and late- type
spirals which have rising LFs).  From Fig.~10 we find a number of
indicative results. First though, we note the rise in
luminosity-density and number-density in the final radial bin.  This
is likely because of the presence of the second D galaxy which lies
0.7 Mpc from the central cD and may represent an infalling sub-group.
Ignoring the bias introduced by this last bin we find that the
luminosity-density of each class falls in a near linear fashion in
log(j) versus radius with gradients of: $-0.68 \pm 0.06$, $-0.32 \pm
0.06$, $-0.30 \pm 0.18$, for E/S0+cD/D, Sabc and Sd/Irr respectively.
Ignoring the cD/D galaxies results in a gradient of $-0.41 \pm 0.08$,
for the E/S0s alone (i.e., consistent with the mid-type
population). This is of course an independent confirmation of the well
known morphology-density relation \citet{dressler97}. Note that the
exclusion of the cD/Ds has little impact upon the derived Schechter
function for early-types, (c.f. dashed line on Fig.~8, middle, E/S0s,
see also Table 5).  Similarly the number-densities also fall near
linearly in log(N) versus radius with a significant variation in
gradient depending on type ($-0.68 \pm 0.21$, $-0.28 \pm 0.08$, $+0.02
\pm 0.07$, for E/S0+cD/Ds, Sabcs and Sd/Irrs
respectively)\footnote{Note that a projected profile with $\rho
\propto r^{-k}$ is roughly equivalent to a real profile of $\rho
\propto r^{-1-k}$, hence the positive projected profile for Sd/Irrs
still implies a decreasing 3D radial profile.}.

From Fig.~10 two clear conclusions can be drawn. Firstly the classical
result that early-type galaxies are more centrally concentrated in
number than mid-type spirals which in turn are more centrally
concentrated than late-type irregulars.  Secondly the flat
number-density profile of late-types implies that the core must be
devoid of late-types which therefore exist exclusively in the cluster
halo, independently confirming the result of \citet{boyce01}.  This
halo extends beyond the field-of-view studied here but from the
luminosity-density profile it is unlikely to contribute significantly
to the total luminosity-density at any radii. Within the field-of-view
studied we note that the total luminosity-density, within all annuli,
is divided into ($72 \pm 13$) per cent E/S0+cD/Ds, ($26 \pm 3$) per
cent Sabcs, and ($2 \pm 1$) per cent Sd/Irrs. This can be compared to
those derived from the SDSS-EDR field LFs shown above (where
$j=\phi^*L^*\Gamma(\alpha+2)$) of 29 per cent, 59 per cent, and 12 per
cent for early-, mid-, and late- types respectively.  Neglecting the
cD/Ds changes the cluster percentages to 63 per cent, 34 per cent and
3 per cent respectively (with similar errors).  

As a comparison we note that values of 33 per cent, 53 per cent, and
14 per cent for the field were derived by \citet{driver99} for a
volume limited sample at $z \approx 0.45$ drawn from the Hubble Deep
Field and classified using the same ANN classifiers as used here. The
consensus between these two independent field studies is reassuring
and provides some indication of the associated errors.  If one assumes
that both the field and cluster environments originate from an
identical shape primordial mass spectrum but with differing
amplitudes, this discrepancy {\it must} be due to an
additional/accelerated evolutionary mechanism(s) over those at work in
the field.  From this data alone one cannot argue factually for the
exact nature of this mechanism other than it has the {\it net} effect
of converting later types towards earlier types, and is most efficient
in the cluster core. In fact if one crudely adopts conservation of
luminosity (strictly more valid at longer wavelengths) then up to 50
per cent of the ellipticals must have been formed from mid- or late-
type spirals. This requires some contrivance given the apparent
universality of the morphological LFs between the field and A868
environment although far more data for both field and clusters are
required before any real significance can be attached to the
difference seen, as well as a more fully consistent classification
scheme.

In general the results here are consistent with the conventional
picture whereby the core environment is hostile to disks and converts
mid-types to early-types --- which remain captured in the core --- and
destroys late-types entirely as they transit through the core.

\section{Conclusions}
We report the first reconstruction of morphological luminosity
functions for a cluster environment since the founding work of
BST. Through the method of background subtraction we recover the
overall LF seen for A868 in a previous ground-based study, but which
used an entirely different region of sky for the background
subtraction, this adds credence to the methodology of background
subtraction for this very rich cluster at least.  In our analysis we
lay down a methodology for accounting for background clustering bias
missing in previous analysis of this type and addressing concerns
raised by \citet{valotto01}.

The overall cluster LF is comparable to the general field LF (2dFGRS)
and we find that the early-, mid- and late- type LFs are all
consistent with the field LFs.  However the errors are such that one
can not yet argue convincingly for, or against, ubiquitous
morphological LFs as proposed by BST.

In exploring the luminosity- and number- density radial profiles we
find flat profiles for late-types and argue that this implies an
absence of late-type galaxies in the core region. Furthermore we find
a significantly skewed luminosity-density breakdown towards
early-types, as compared to the field.  We {\it speculate} that this
implies that cluster cores are in essence disk-destroying engines
resulting in the build up of a hot inter-cluster member and the
formation of a tightly bound population of intermediate-luminosity
core ellipticals most likely formed from mid-type bulges.

Finally from our error analysis we note that more definitive results
can only be obtained from the combination of cluster data, as the
dominant error at bright magnitudes is simply the number of cluster
members.  Such data is is now becoming freely available via the {\it
Sloan Digital Sky Survey} and the {\it Hubble Space Telescope}
Archives. In this paper we have laid down a rigorous methodology for
the analysis of such data and look forward to an illuminating era.

\acknowledgements We would like to thank Ray Lucas for help with the
planning of the observations and Roberto De Propris and an anonymous
referee for helpful comments on the manuscript. Support for program
\#8203 was provided by NASA through a grant from the Space Telescope
Science Institute, which is operated by the Association of
Universities for Research in Astronomy, Inc., under NASA contract NAS
5-26555.

\clearpage

\figcaption[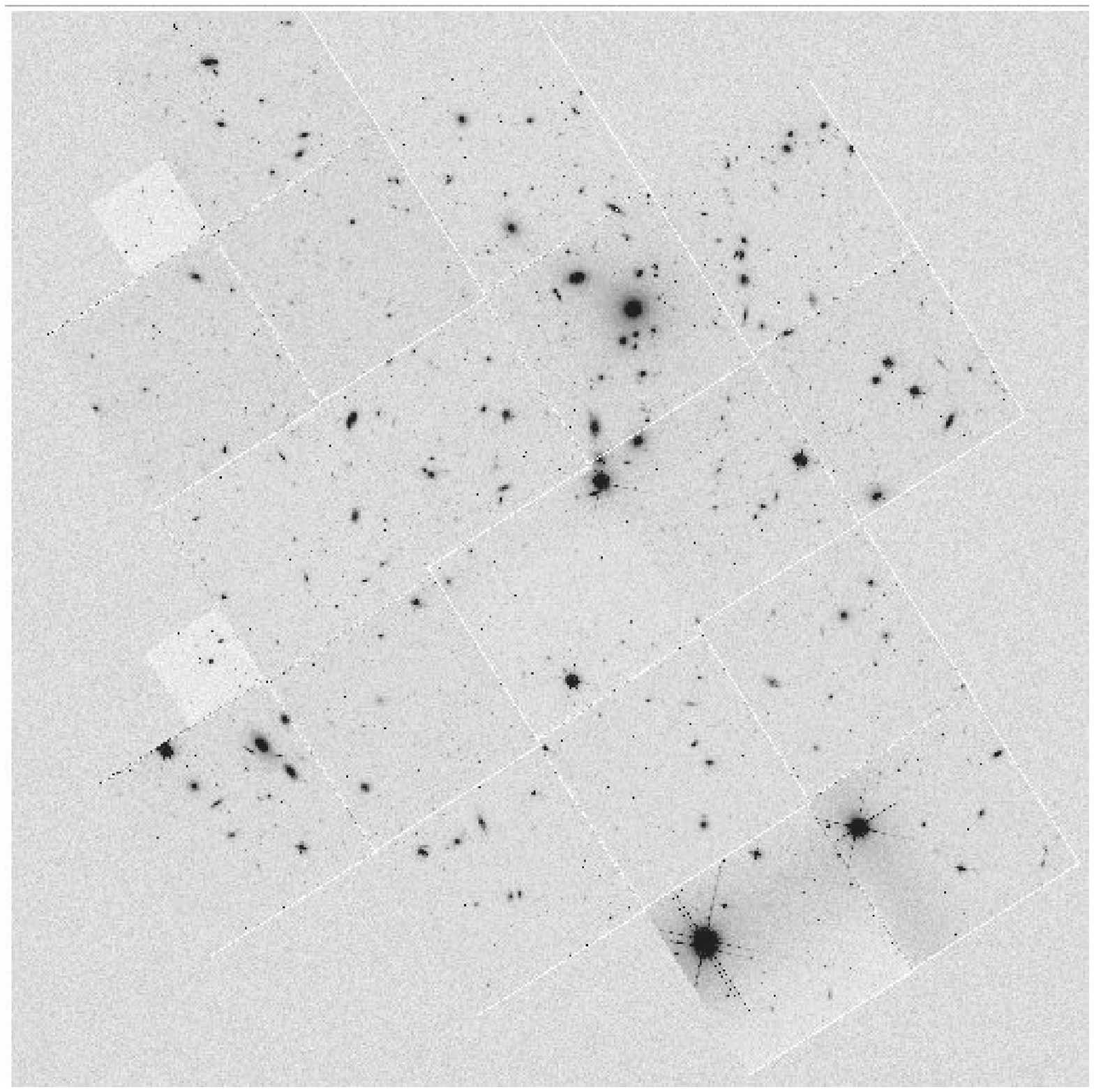]{The full six pointing mosaic of the A868
cluster and environs.  North is up with East to the left. The box is
6.62 arcminutes (or 1.08 Mpc at $z =0 .158$) on each side.  The
cluster core is clearly visible and shown in more detail in Fig.\~2.}
\label{fig1}

\figcaption[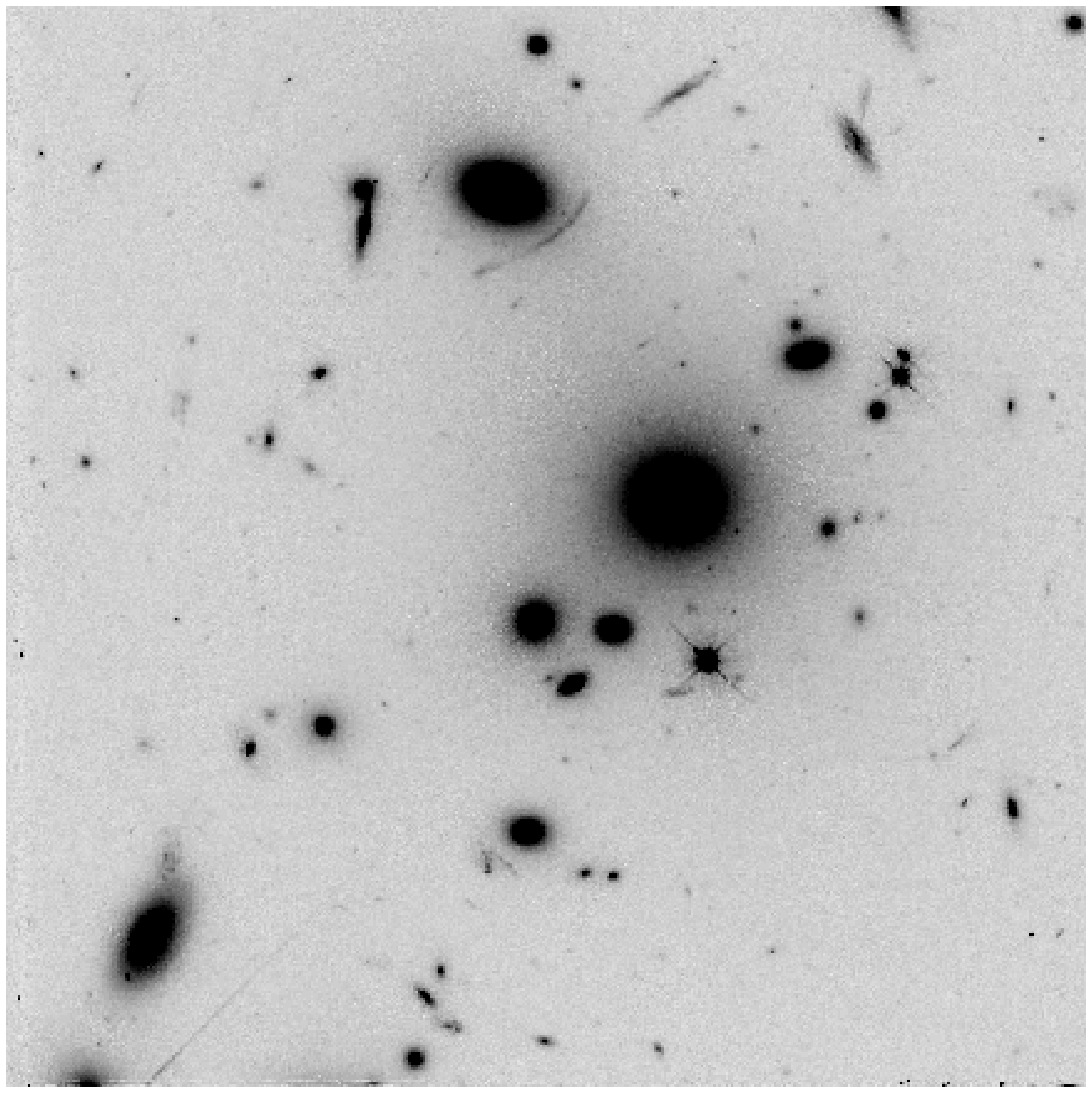]{A single WFPC2 chip showing the core of the
rich cluster A868. The image is approximately 1.5 arcminutes on each
side (0.25 Mpc). Clearly visible is the central cD and one of the D
galaxies with numerous examples of gravitational lenses.}
\label{fig2}

\figcaption[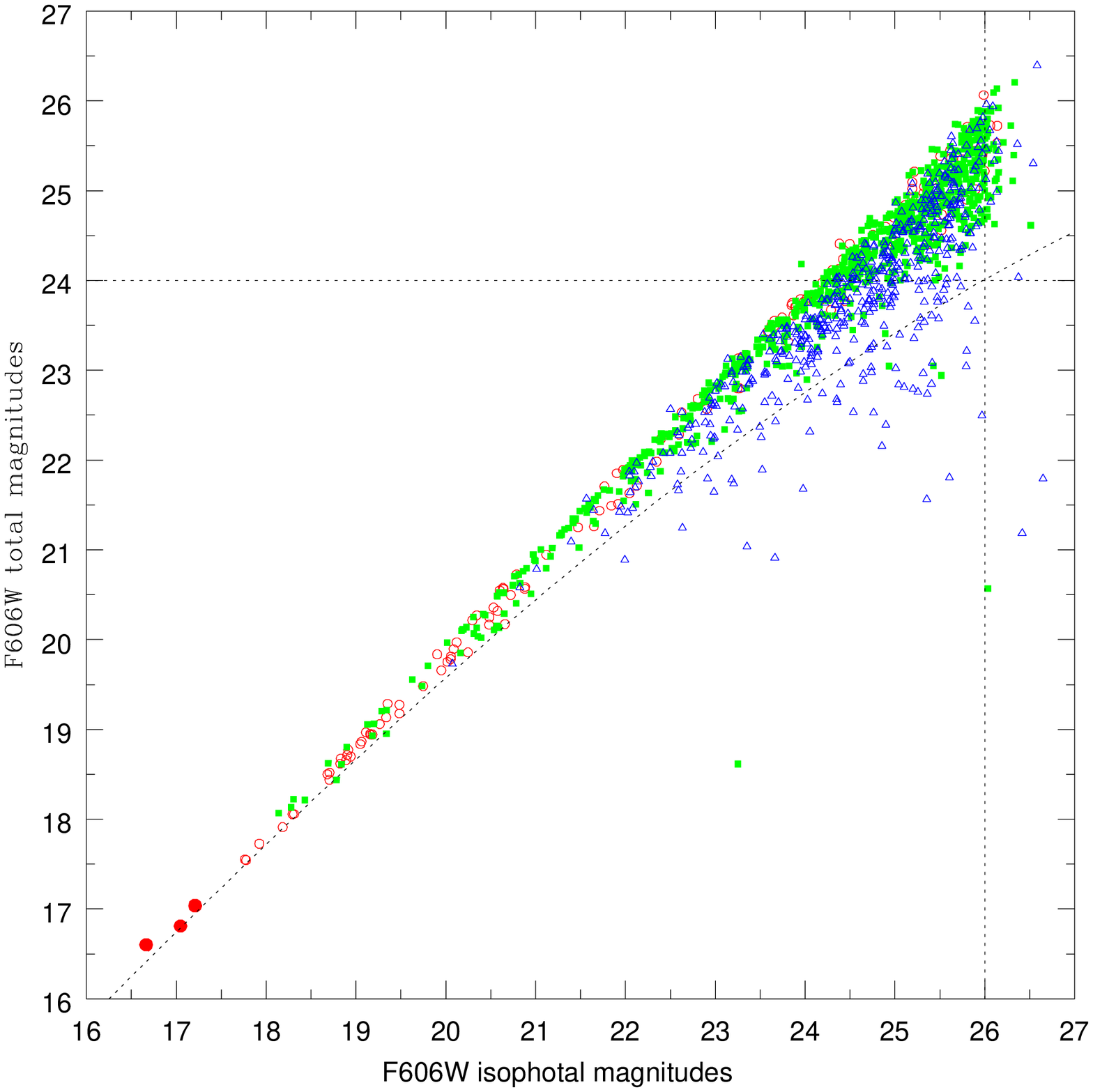]{Total versus isophotal magnitudes for the
full A868 mosaic. The cD/Ds, early-, mid- and late-types are denoted
as filled circles, open circles, filled squares and open triangles
respectively. The dotted line shows the adopted cap to the isophotal
corrections. (Note that the one spiral furtherest from the unity line
lies close to a bright elliptical resulting in the extreme isophotal
correction.)}
\label{fig3}

\figcaption[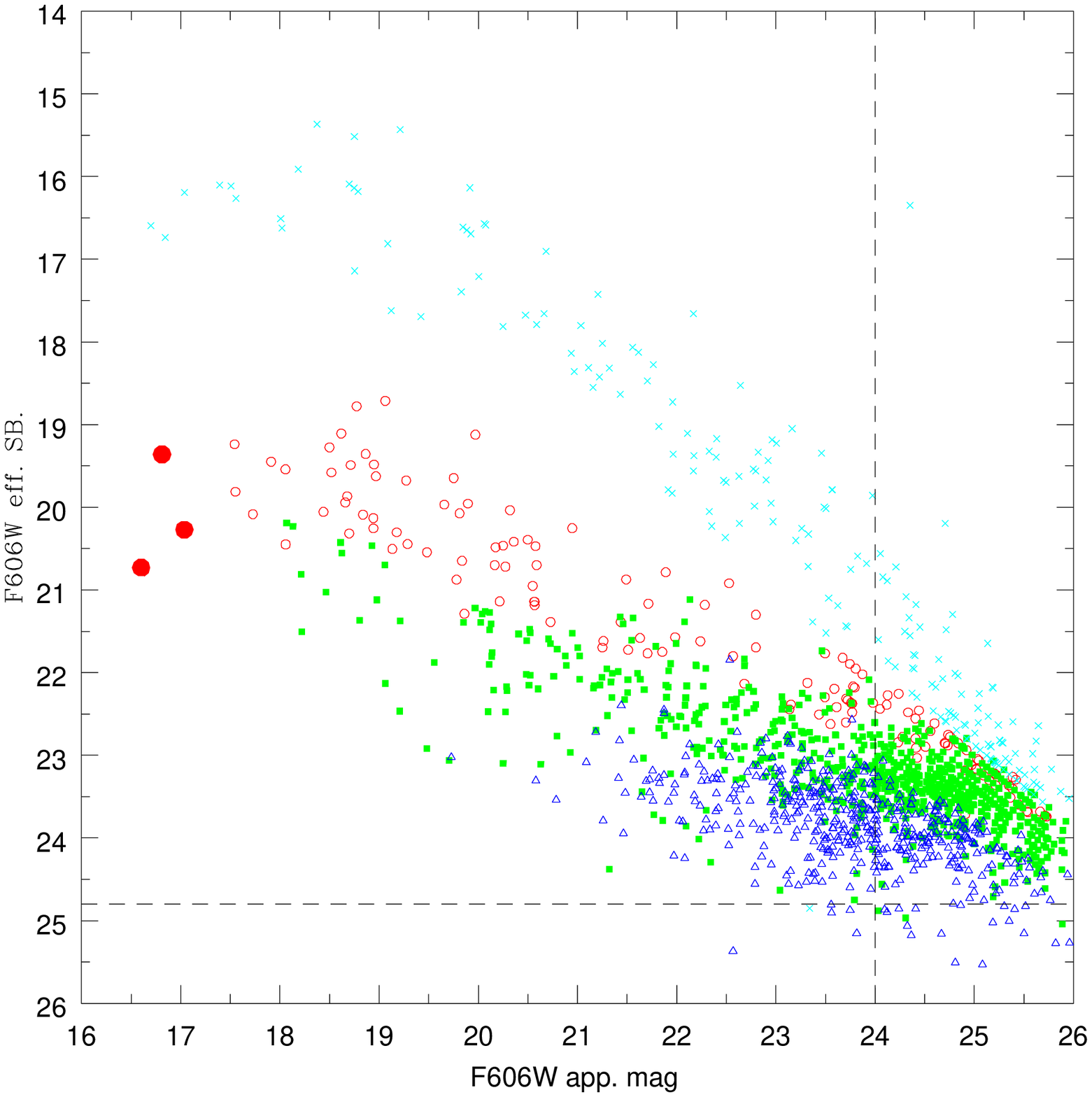]{The apparent bivariate brightness
distribution for galaxies in the A868 sight-line. The effective
surface brightness is derived from the measured major-axis half-light
radius ($\mu_{eff} = m+2.5 \log_{10}(2 \pi r^{2}_{hlr})$). Large
filled circles denote cD/Ds, open circles early-types, filled squares
mid-types, triangles late-types and crosses stars. The dashed lines
denote the limiting surface brightness and star-galaxy separation
limit.}
\label{fig4}

\figcaption[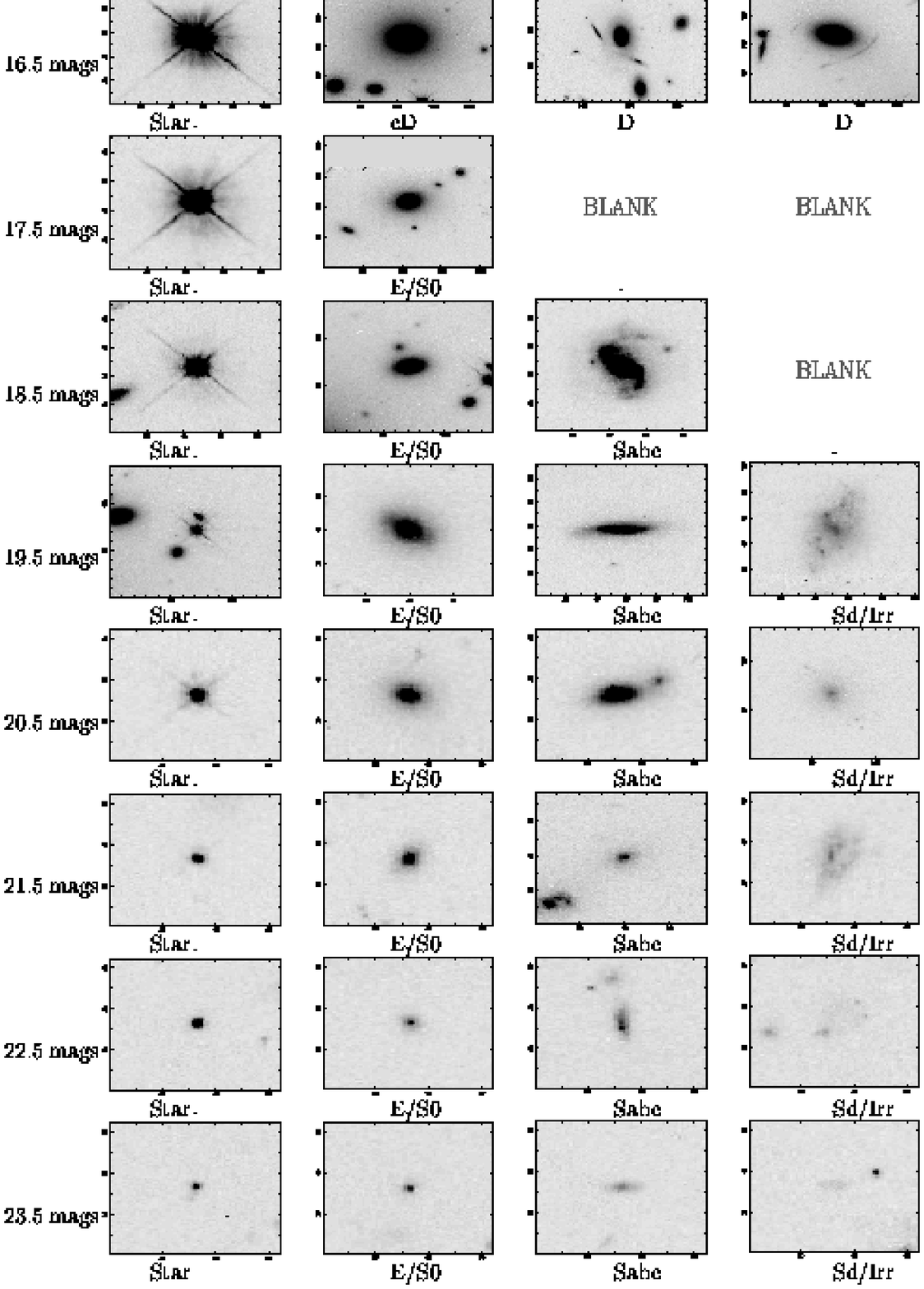]{A random sample of stars, early-, mid- and
late- types (across) versus apparent magnitude (down).}
\label{fig5}

\figcaption[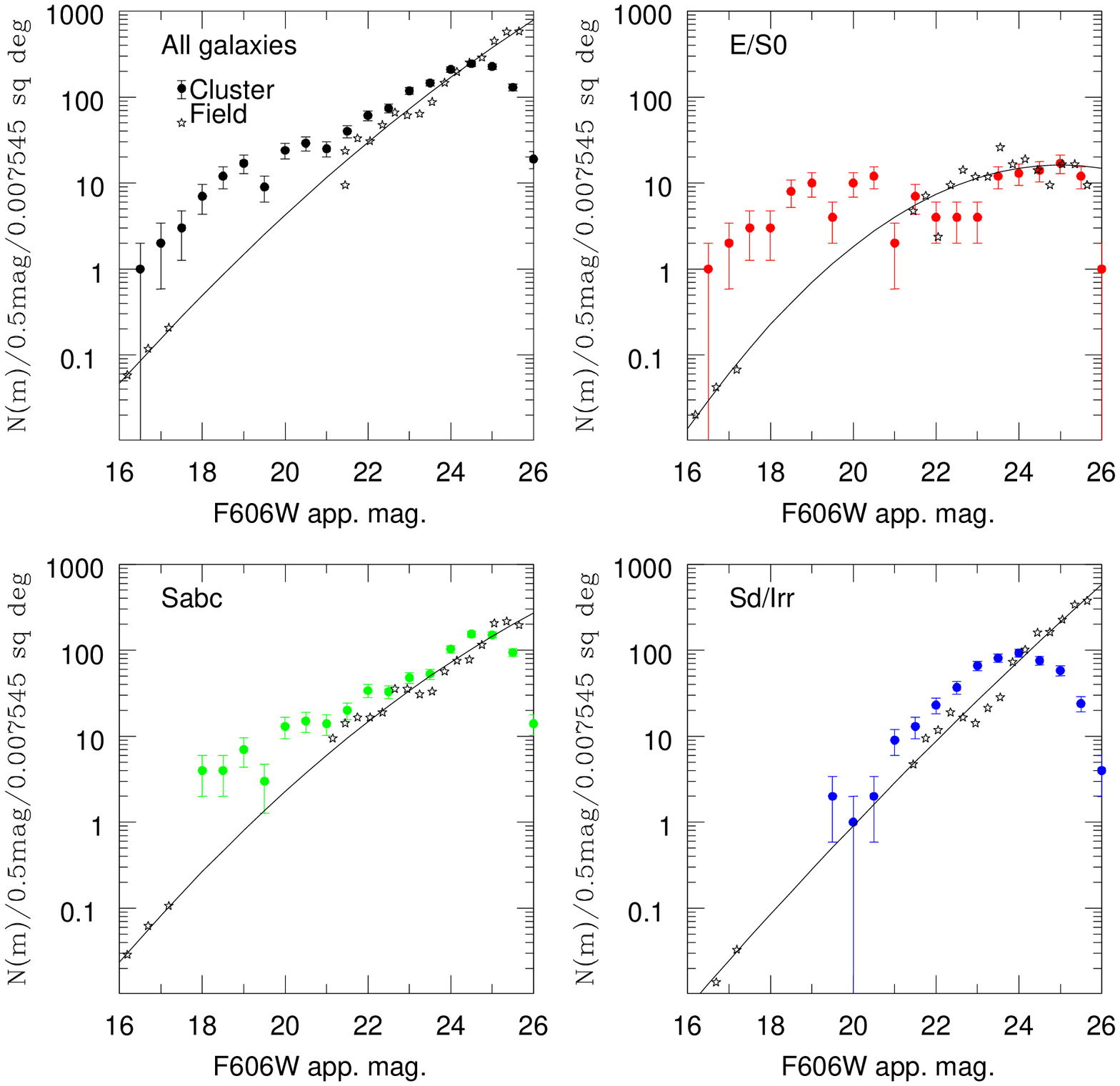]{Cluster sight-line and reference field counts
(scaled to the A868 field-of-view size of 0.007545 sq degrees). Counts
are shown for all galaxies, early-, mid- and late-types. The
solid lines shows a 2nd order polynomial fit to the field count data. Errors
are purely Poisson at this point.}
\label{fig6}

\figcaption[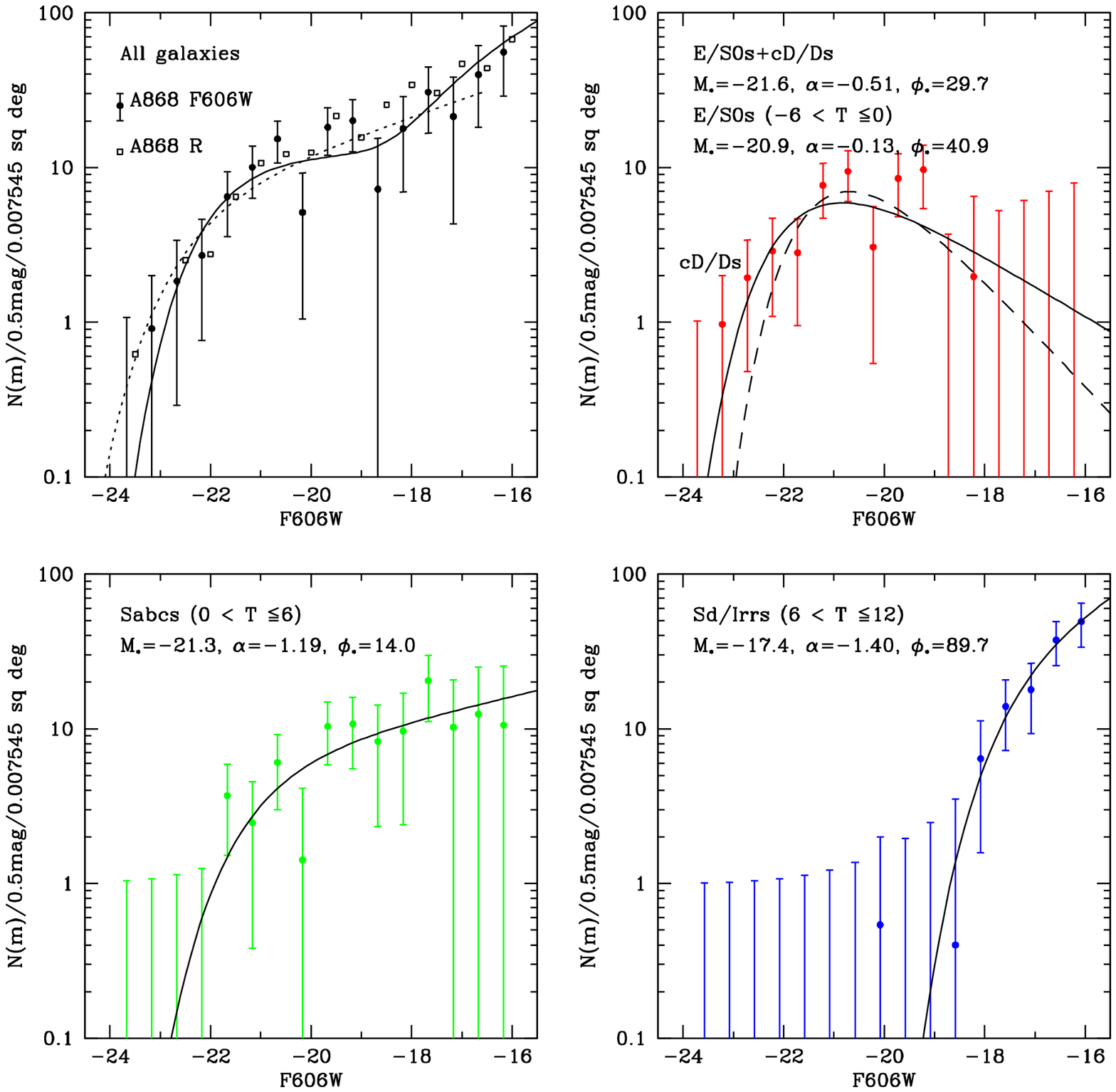]{The recovered luminosity distributions for the
cluster population after subtracting the reference field counts from
the A868 sight-line counts. The errorbars now include the full error
analysis (i.e., five error components including 3 Poisson and 2
clustering errors). The solid lines show the $\chi^2$-minimised
Schechter function fits to the data. In the case of all galaxies we
also show the 2dFGRS composite cluster luminosity function
(\citep{dp03}) transposed to the F606W filter and renormalized to
match the data. The squares show the previous deeper $R$-band cluster
results from \citet{driver98a}.}
\label{fig7}

\figcaption[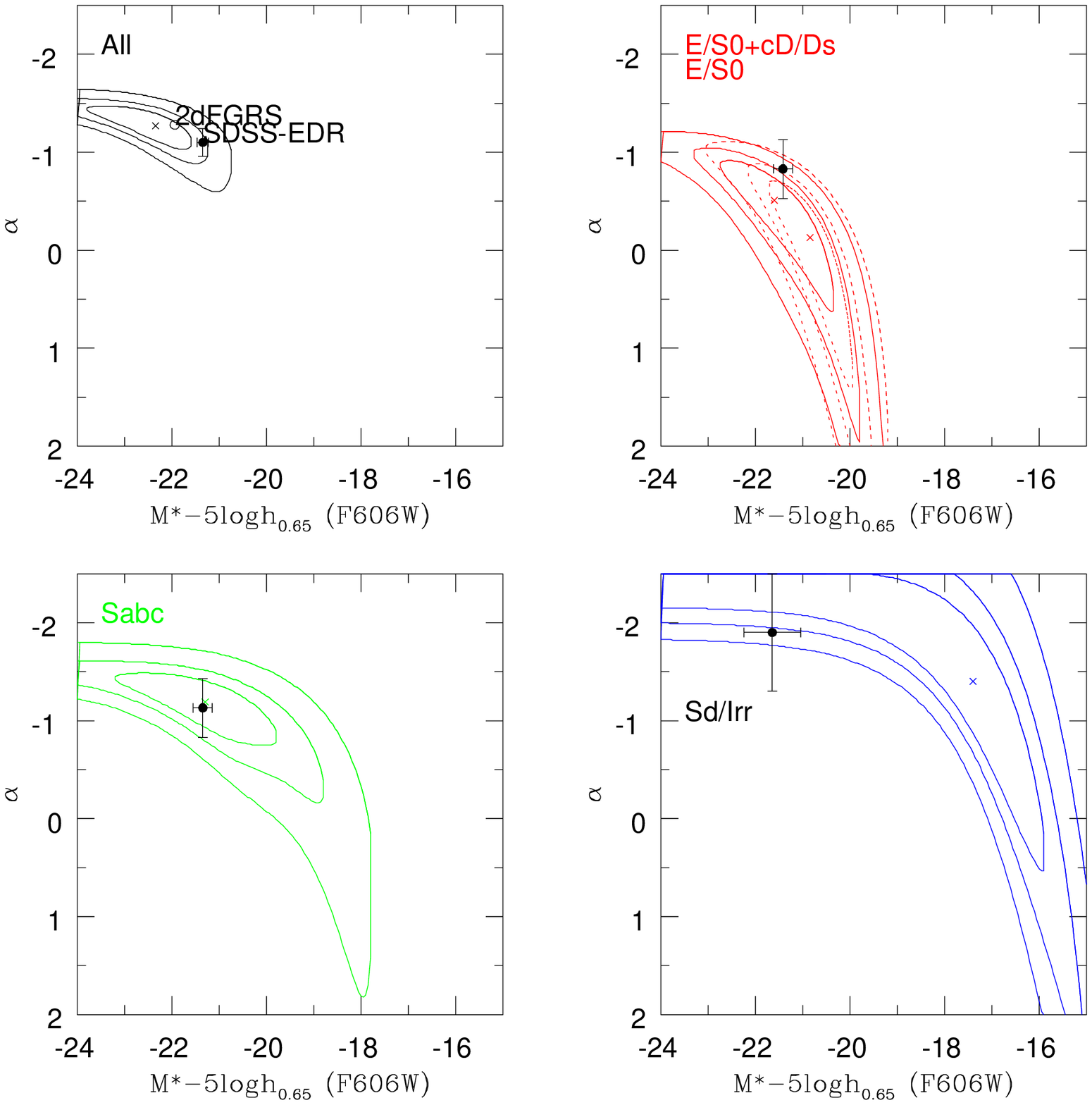]{The 1, 2 and 3 $\sigma$ error contours for
the Schechter function fits shown in Fig.\~9. The crosses shows the
actual best fit points. The solid points with errorbars show the
recent field estimates based on SDSS-EDR data by Nakamura et al
2003. The open data point shows the recent composite cluster LF
estimated from the 2dFGRS \citep{dp03}, errors are comparable to the
symbol size. For the ellipticals the solid contours show the fit to
E/S0s+cD/Ds and the dotted contours the fits to E/S0s only.}
\label{fig8}

\figcaption[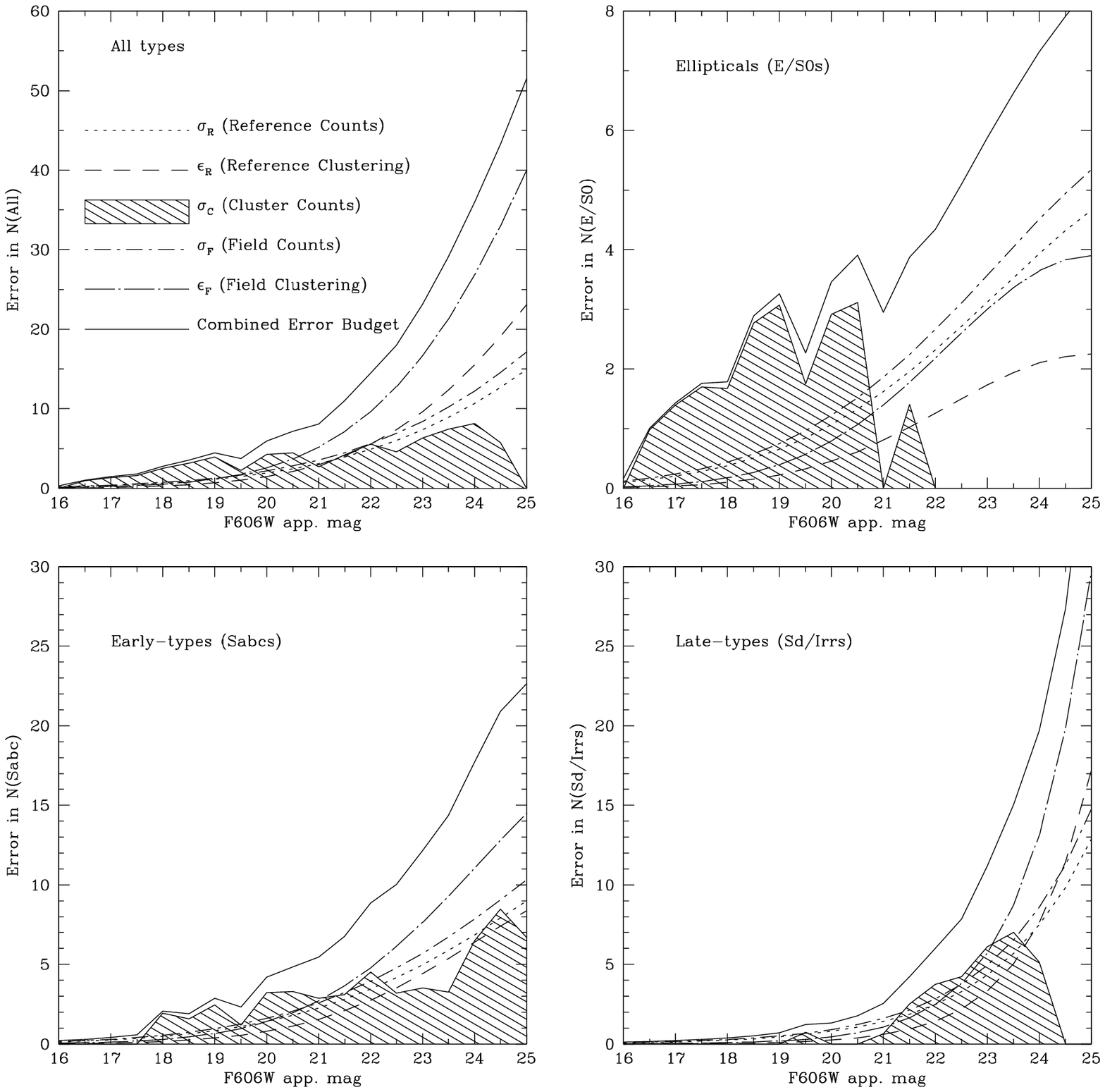]{The five error components to the cluster
luminosity distributions for all galaxies, ellipticals, early-type and
late-type galaxies. In most cases the Poisson error in the cluster
population dominates at brighter magnitudes and the clustering error
in the field population dominates at faint magnitudes.}
\label{fig9}

\figcaption[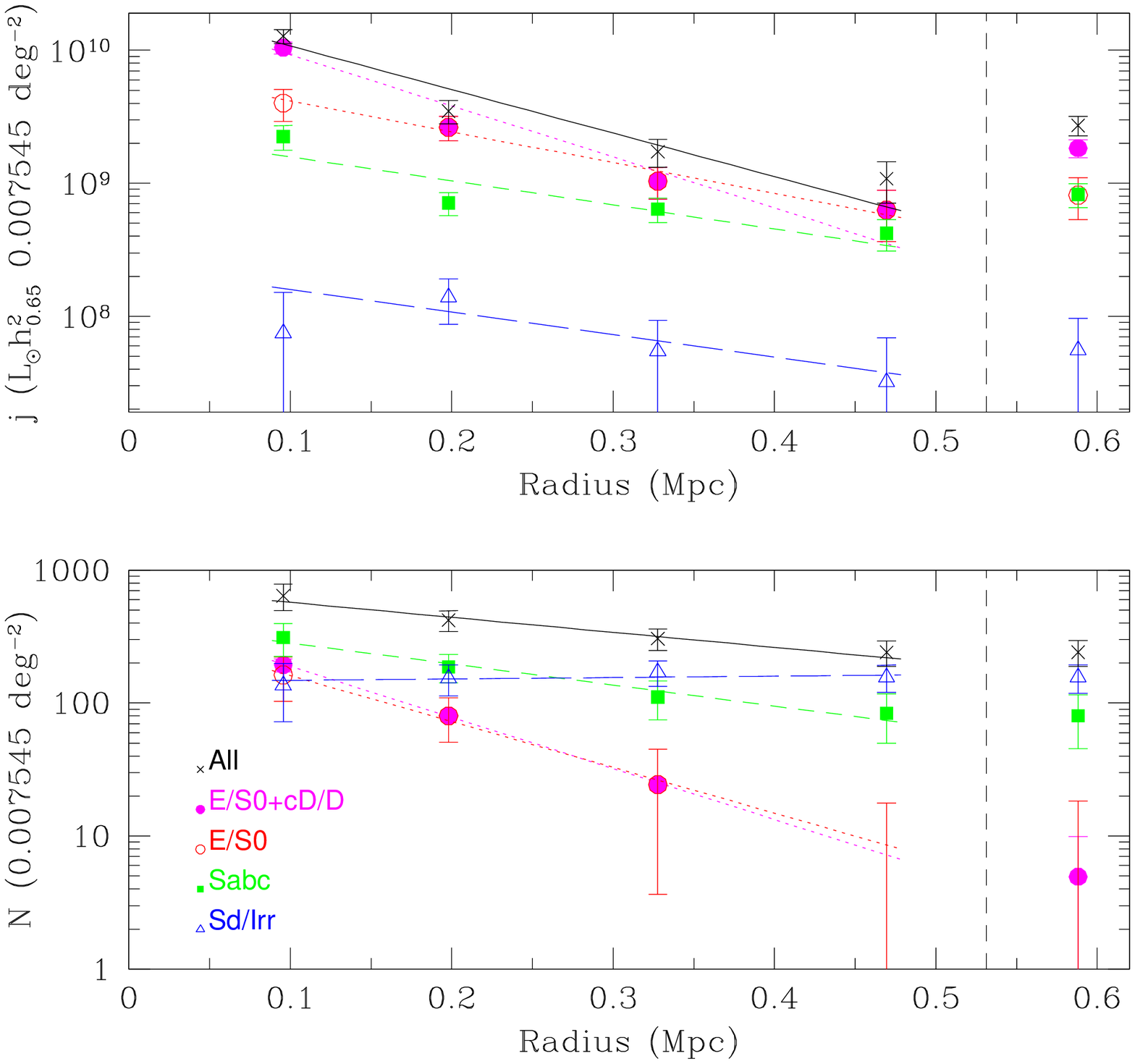]{(upper) The luminosity-density profiles derived
from the absolute magnitude range $-24 < M_{F606w} < -16$ for all
(crosses, solid), cD/D+E/S0 (pentagons, dotted line), E/S0 (circles,
dotted line), Sabc (squares, short dashed), Sd/Irr (triangles, long
dashed line) in arbitrary units in five annuli centered around the
central cD galaxies.  (lower) The equivalent number-density profiles
labeled as above.}.
\label{fig10}

\clearpage

\begin{deluxetable}{lrrrr}
\tablecaption{Summary of morphological classification overrides}
\label{table1} \tablewidth{0pt} \tablehead{
\colhead{} &
\colhead{Ellipticals} & 
\colhead{Early-types} &
\colhead{Late-types} &
\colhead{Stars}}
\startdata
Ellipticals & - & 4 & 0 & 2 \\
Early-types & 2 & - & 20 & 0 \\
Late-types & 1 & 32 & - & 0\\
Stars & 2 & 3 & 1 & -\\
Junk & 2 & 11 & 2 & 6\\
\enddata
\end{deluxetable}

\begin{deluxetable}{lllll}
\tablecaption{Number Count data for all, early-, mid- and late- type
reference field galaxies per 0.007545 sq degrees. The errors include
both Poisson and clustering components}
\label{table2} \tablewidth{0pt} \tablehead{
\colhead{mag} & 
\colhead{$N(All)$} &
\colhead{$N(E/S0)$} &
\colhead{$\Delta N(Sabc)$} &
\colhead{$\Delta N(Sd/Irr)$}}
\startdata
$15.190$&$0.016+/-0.003$&$0.003+/-0.001$&$0.010+/-0.002$&$0.003+/-0.001$\\
$15.690$&$0.030+/-0.005$&$0.011+/-0.002$&$0.012+/-0.003$&$0.006+/-0.002$\\
$16.190$&$0.057+/-0.009$&$0.020+/-0.004$&$0.028+/-0.005$&$0.009+/-0.002$\\
$16.690$&$0.115+/-0.015$&$0.041+/-0.006$&$0.060+/-0.008$&$0.013+/-0.002$\\
$17.190$&$0.202+/-0.023$&$0.066+/-0.008$&$0.104+/-0.012$&$0.032+/-0.004$\\
$21.450$&$23.003+/-8.751$&$4.601+/-3.395$&$9.201+/-5.061$&$4.601+/-3.395$\\
$21.750$&$32.204+/-10.655$&$6.901+/-4.205$&$13.802+/-6.346$&$9.201+/-4.938$\\
$22.050$&$29.904+/-9.886$&$2.300+/-2.337$&$16.102+/-6.849$&$11.502+/-5.544$\\
$22.350$&$46.006+/-12.811$&$9.201+/-4.847$&$16.102+/-6.740$&$18.402+/-7.187$\\
$22.650$&$64.409+/-15.664$&$13.802+/-6.018$&$18.402+/-7.187$&$16.102+/-6.566$\\
$22.950$&$59.808+/-14.452$&$11.502+/-5.394$&$34.505+/-10.357$&$13.802+/-5.962$\\
$23.250$&$62.108+/-14.432$&$11.502+/-5.357$&$34.505+/-10.154$&$20.703+/-7.409$\\
$23.550$&$85.111+/-17.329$&$25.303+/-8.212$&$29.904+/-9.162$&$27.604+/-8.631$\\
$23.850$&$142.619+/-24.035$&$16.102+/-6.342$&$32.204+/-9.436$&$71.309+/-15.048$\\
$24.150$&$190.925+/-28.632$&$18.402+/-6.773$&$55.207+/-12.822$&$98.913+/-18.157$\\
$24.450$&$246.133+/-33.231$&$13.802+/-5.783$&$73.610+/-15.030$&$156.421+/-24.024$\\
$24.750$&$280.637+/-35.223$&$9.201+/-4.670$&$75.910+/-15.027$&$158.721+/-23.569$\\
$25.050$&$437.058+/-47.259$&$16.102+/-6.221$&$112.715+/-18.849$&$220.829+/-28.661$\\
$25.350$&$558.974+/-54.724$&$16.102+/-6.201$&$200.127+/-26.792$&$331.244+/-36.906$\\
$25.650$&$565.875+/-52.836$&$9.201+/-4.643$&$211.628+/-27.051$&$365.748+/-38.259$\\
\enddata		
\end{deluxetable}

\begin{deluxetable}{cl}
\tablecaption{2rd order polynomial fits to the field number-count data over the range $15 < m < 24.25$}
\label{table3} \tablewidth{0pt} \tablehead{
\colhead{Fit}&
\colhead{$\chi^2/\nu$}}
\startdata
$log_{10}N(All)dm = -14.752 + 1.103 m_{F606W} - 0.0166 m_{F606W}^2$ & 7.8/12 \\
$log_{10}N(E/S0)dm = -18.595 + 1.484 m_{F606W} - 0.0273 m_{F606W}^2$ & 10.0/11 \\
$log_{10}N(Sabc)dm = -16.770 + 1.290 m_{F606W} - 0.0215 m_{F606W}^2$ & 10.0/12 \\
$log_{10}N(Sd/Irr)dm = -11.491 + 0.639 m_{F606W} - 0.0035 m_{F606W}^2$ & 14.2/11 \\
\enddata
\end{deluxetable}

\begin{deluxetable}{lllll}
\tablecaption{The estimated cluster population for all, early-, mid-,
and late- type A868 cluster galaxies per 0.007545 sq degrees. The
errors include both Poisson and clustering components}
\label{table4} \tablewidth{0pt} \tablehead{
\colhead{mag} & 
\colhead{$N(All)$} &
\colhead{$N(E/S0)$} &
\colhead{$N(Sabc)$} &
\colhead{$N(Sd/Irr)$}}
\startdata
$16.00$&$-0.05+/-1.07$&$-0.01+/-1.02$&$-0.02+/-1.04$&$-0.01+/-1.01$\\
$16.50$&$0.91+/-1.10$&$0.97+/-1.03$&$-0.05+/-1.07$&$-0.01+/-1.02$\\
$17.00$&$1.84+/-1.55$&$1.94+/-1.46$&$-0.09+/-1.14$&$-0.02+/-1.04$\\
$17.50$&$2.70+/-1.94$&$2.89+/-1.80$&$-0.16+/-1.25$&$-0.04+/-1.07$\\
$18.00$&$6.46+/-2.90$&$2.81+/-1.86$&$3.70+/-2.18$&$-0.08+/-1.13$\\
$18.50$&$10.03+/-3.69$&$7.67+/-2.97$&$2.47+/-2.09$&$-0.14+/-1.22$\\
$19.00$&$15.32+/-4.67$&$9.43+/-3.38$&$6.07+/-3.07$&$-0.26+/-1.37$\\
$19.50$&$5.13+/-4.08$&$3.06+/-2.52$&$1.42+/-2.69$&$0.54+/-1.46$\\
$20.00$&$18.19+/-6.20$&$8.51+/-3.72$&$10.38+/-4.51$&$-0.83+/-1.96$\\
$20.50$&$20.10+/-7.43$&$9.69+/-4.25$&$10.75+/-5.23$&$-0.47+/-2.48$\\
$21.00$&$7.25+/-8.20$&$-1.46+/-3.71$&$8.29+/-5.95$&$0.40+/-3.12$\\
$21.50$&$17.84+/-10.89$&$1.97+/-4.55$&$9.67+/-7.26$&$6.41+/-4.83$\\
$22.00$&$30.70+/-14.03$&$-3.09+/-5.25$&$20.47+/-9.34$&$13.94+/-6.69$\\
$22.50$&$21.34+/-17.02$&$-5.67+/-6.12$&$10.23+/-10.52$&$17.90+/-8.61$\\
$23.00$&$39.80+/-21.54$&$-8.79+/-7.02$&$12.44+/-12.63$&$37.43+/-11.88$\\
$23.50$&$55.65+/-26.65$&$-4.38+/-7.92$&$10.57+/-14.81$&$49.35+/-15.55$\\
$24.00$&$66.95+/-32.43$&$-7.34+/-8.78$&$42.17+/-18.15$&$26.27+/-19.69$\\
\enddata		
\end{deluxetable}

\begin{deluxetable}{lccccc}
\tablecaption{Derived Schechter function parameters for the overall
and morphological luminosity distributions of the Rich cluster Abell
868.} \label{table5} \tablewidth{0pt} \tablehead{
\colhead{Morphological} & \colhead{T-type} &
\colhead{$\phi_{*}h^3_{0.65}$} & \colhead{$M_{F606W}^*-5log_{10}h_{0.65}$}
& \colhead{$\alpha$} & \colhead{$\chi^2/\nu$} \\ \colhead{Class} &
\colhead{Range} & \colhead{($0.007545$ sq deg$^{-1}$)} &
\colhead{(mag)} & \colhead{} & \colhead{} }

\startdata 
All       & $-6 < T < 9$    & 16.4 & $-22.4^{+0.6}_{-0.6}$ & $-1.27^{+0.13}_{-0.15}$ & 8.8/12 \\ 
E/S0+cD/D & $-6 < T \leq 0$ & 29.2 & $-21.6^{+0.6}_{-0.5}$ & $-0.51^{+0.2}_{-0.3}$ & 7.9/7 \\ 
E/S0      & $-6 < T \leq 0$ & 41.2 & $-20.9^{+0.4}_{-0.4}$ & $-0.13^{+0.4}_{-0.4}$ & 6.9/5 \\ 
Sabc      & $0 < T \leq 6$  & 14.0 & $-21.3^{+1.0}_{-0.9}$ & $-1.19^{+0.2}_{-0.2}$ & 6.0/10 \\ 
Sd/Irr    & $6 < T \leq 9$  & 89.7 & $-17.4^{+0.7}_{-0.7}$ & $-1.40^{+0.6}_{-0.5}$ & 0.7/4 \\ 
\enddata
\end{deluxetable}

\clearpage

\plotone{driver.fig1_small.ps}

\plotone{driver.fig2_small.ps}

\plotone{driver.fig3.ps}

\plotone{driver.fig4.ps}

\plotone{driver.fig5_small.ps}

\plotone{driver.fig6.ps}

\plotone{driver.fig7.ps}

\plotone{driver.fig8.ps}

\plotone{driver.fig9.ps}

\plotone{driver.fig10.ps}

\end{document}